\documentclass[aip,reprint]{revtex4-1}

\usepackage{graphicx}
\usepackage[normalem]{ulem}
\usepackage[usenames]{color}

\begin{document}


\title{Fabrication and Electrical Characterization of Fully CMOS Si Single Electron Devices}


\author{P.~J. Koppinen}
\affiliation{National Institute of Standards and Technology (NIST), Gaithersburg, MD 20889, USA} 
\affiliation{Joint Quantum Institute, University of Maryland, College Park, MD 20742, USA}
\author{M.~D. Stewart, Jr.}
\email[Author to whom correspondence should be addressed. Electronic mail: ]{michael.d.stewart@nist.gov}
\affiliation{National Institute of Standards and Technology (NIST), Gaithersburg, MD 20889, USA} 
\affiliation{Joint Quantum Institute, University of Maryland, College Park, MD 20742, USA}
\author{Neil M. Zimmerman}
\affiliation{National Institute of Standards and Technology (NIST), Gaithersburg, MD 20889, USA} 


\date{\today}

\begin{abstract}
  We present electrical data of silicon single electron devices fabricated with CMOS techniques and protocols. The easily tuned devices show clean Coulomb diamonds at $T=30$ mK and charge offset drift of 0.01 e over eight days. In addition, the devices exhibit robust transistor characteristics including uniformity within about 0.5 V in the threshold voltage, gate resistances greater than 10 G$\Omega$, and immunity to dielectric breakdown in electric fields as high as $4$ MV/cm. These results highlight the benefits in device performance of a fully CMOS process for single electron device fabrication.
\end{abstract}
\pacs{73.23.-b, 73.23.Hk, 73.40.Gk, 73.40.Ty, 73.63.-b, 73.63.Kv}

\maketitle 
\section{Introduction and motivation}
Single electron tunneling (SET) devices\cite{SETbook} are promising candidates for a wide variety of nanoelectronics applications, such as sensitive electrometers\cite{Schoelkopf}, thermometers\cite{CBT}, electron pumps and turnstiles for current standards\cite{gallop05,zimmerman03}, and quantum bits for quantum information processing\cite{fujisawa06a,hanson07a,morton11}. 
In recent years, silicon has drawn a lot of attention as a candidate for practical SET devices for several reasons. These advantages include compatibility with complementary metal--oxide--semiconductor (CMOS) processing, good electrostatic control of the tunnel barriers\cite{fujiwara06}, greater device stability as demonstrated by a lack of charge offset drift\cite{NeilJAP,zimmerman07a,hourdakis08a}, and a relative lack of nuclear spins, an important source of decoherence in spin-based quantum information applications\cite{Witzel10}. 

However, to become truly viable in any of these applications, devices must be fabricated which overcome the device to device variations and low yield associated with the single device processing typical of small--scale research programs. Although, at the single device level, the gate voltage variation from one device to another may not be an important parameter, uniform device operation becomes crucial when trying to operate several SET devices simultaneously, e.g., in the large scale integration of SET devices. The choice of device architecture can also impact the integrability of devices. For example, gate to gate variations in an architecture where more than one gate\cite{hanson07a,tracy10} controls a single tunnel barrier can make finding the desired operating point a laborious iterative process.  

In this paper we demonstrate robust behavior and good unformity of easily-tuned, fully CMOS single electron devices, which contain only silicon, thermally--grown silicon dioxide (SiO$_2$) and phorphorous--doped polycrystalline silicon (poly--Si) in the active device region. The motivation for a fully CMOS approach to fabrication is twofold: 1) to minimize the number of impurities and defects near the active device region and 2) to avoid the instabilities associated with metallic oxides and, in particular, aluminum oxide. In this way, we avail ourselves of the best opportunity to fabricate uniform, robust devices. Below, we will discuss and demonstrate the robustness of our devices with respect to basic metal--oxide--semiconductor field--effect transistor (MOSFET) characteristics and SET device operation. In particular, we show that these devices exhibit only small variations of the threshold voltage from device to device, dielectrics which are robust against breakdown, and charge offset stability of the order of 0.01 e over a period of several days. 

\section{Operating principle of the device and fabrication}
\begin{figure}[ht!]
  \includegraphics[width=\columnwidth]{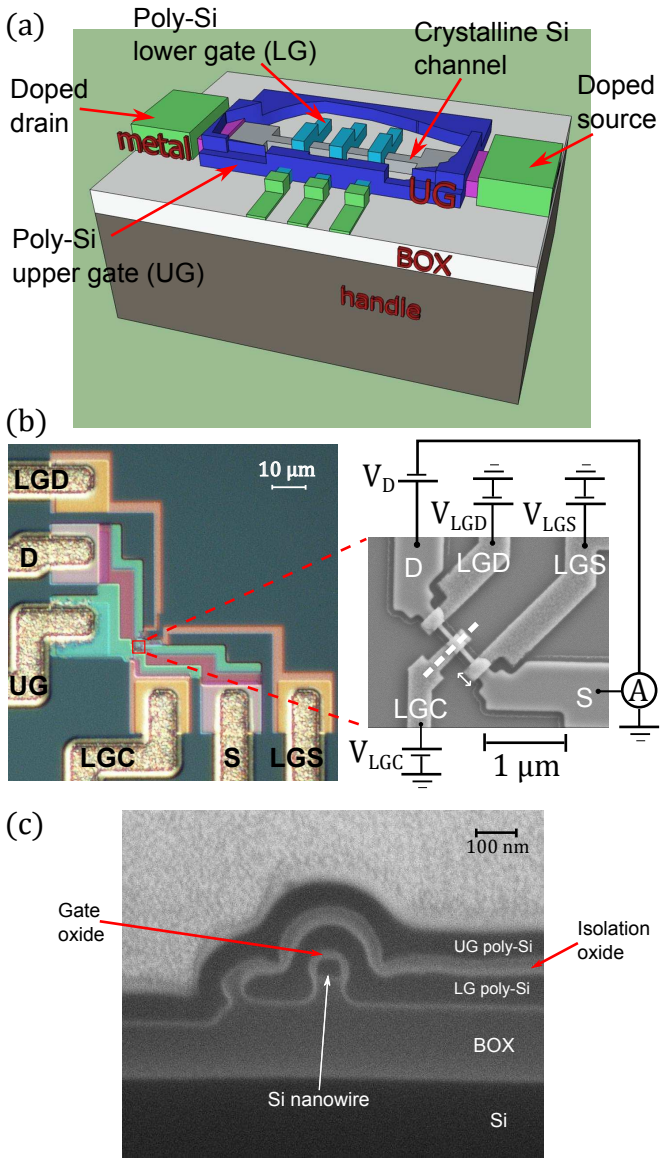}
  \caption{\label{sample}(Color online) (a) A schematic view of a sample. Device operation is described in the text. (b) Left: An optical micrograph of a sample. Right: An SEM micrograph (before upper gate deposition) of the active device area, and schematics of an electrical measurement circuit (does not show V$_{UG}$). Lower gates LGS, LGC and LGD are poly--Si and the conducting channel (S/D) is single crystal Si. Channel and lower gates sit on top of the buried silicon oxide (BOX). The white arrow next to the finger gate indicates the dimension we call gate length. (c) A cross--sectional SEM image of a device along the dashed white line in (b). The darker areas are Si, the gray areas are SiO$_{2}$ and the bright layer on top is a protective layer of Pt deposited prior to the FIB cut. 
}
\end{figure}

\begin{figure}[ht]
  \includegraphics[width=\columnwidth]{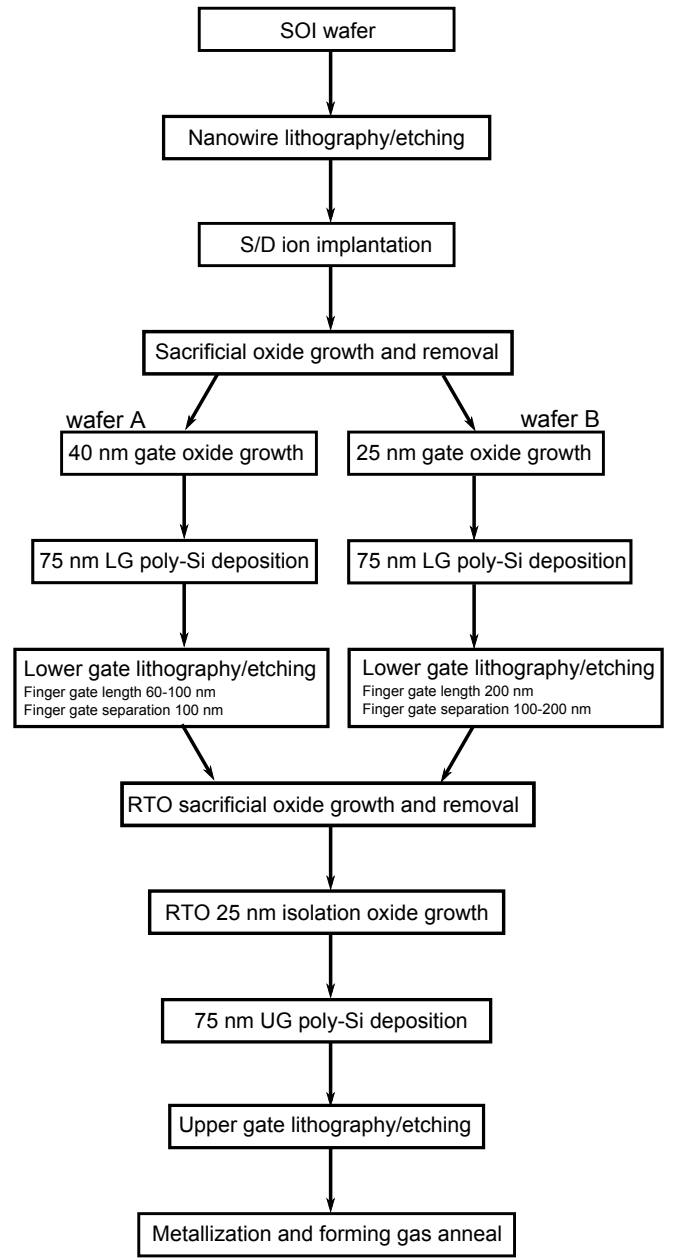}
  \caption{\label{fabflowchart} Flow chart of the condensed fabrication process described in the text.}
\end{figure}

\begin{figure*}[ht!]
  \includegraphics[width=\textwidth]{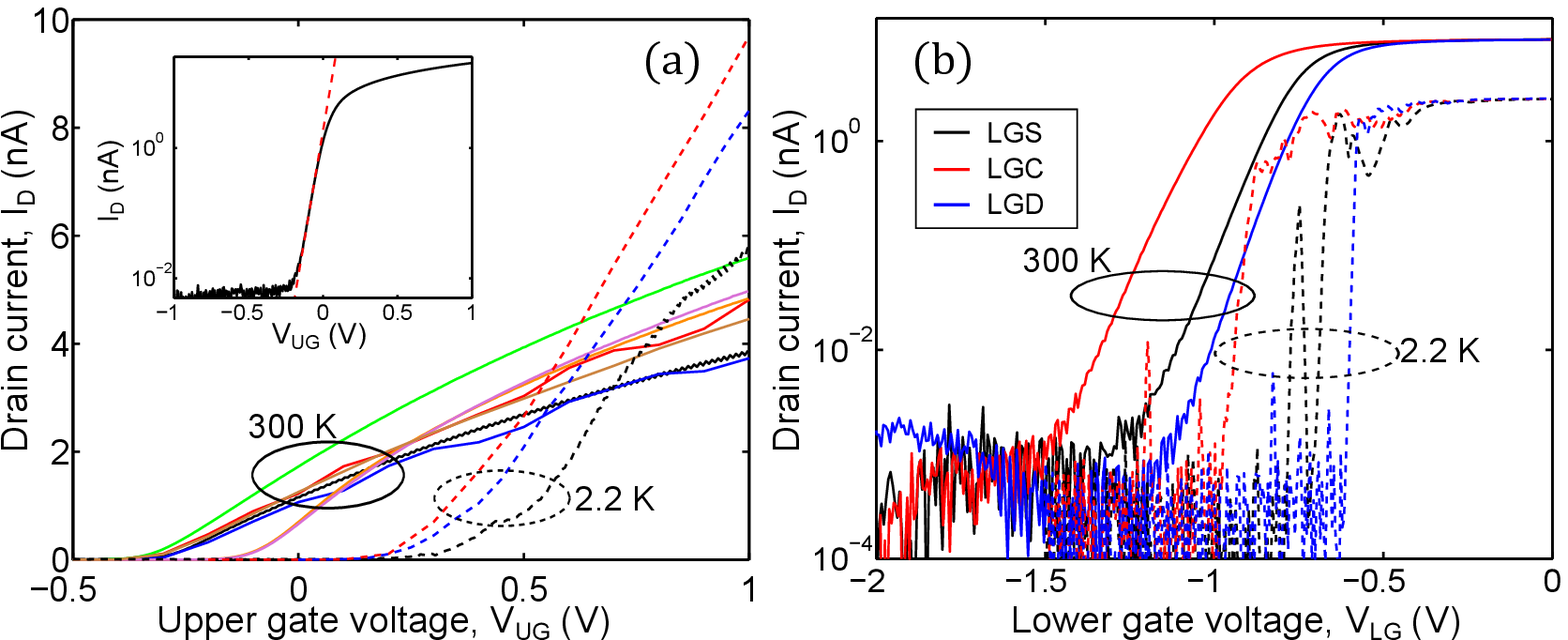}
  \caption{\label{3Gchar}(Color online) (a) Turn--on characteristics of SET devices measured as MOSFETs. Solid lines represent room temperature data for different devices, dashed lines are three different devices measured at 2.2 K. Inset: turn--on characteristic of a single device in semilogarithmic plot measured at room temperature. Devices show good uniformity in threshold voltages and the inset shows sub--threshold slope of 80 mV/decade (slope of red dashed line). (b) Turn--off characteristics of each individual lower gate of a device measured at room temperature (solid lines) and at 2.2 K (dashed lines). Black represents LGS, red (gray, the leftmost curve at both temperatures) LGC and blue (gray, the rightmost curve at both temperatures) LGD. Upper gate voltage, V$_{UG}$ was 1.3 V and 2 V for the room temperature data and 2.2 K data, respectively. All lower gates turn off conduction in the channel, and show about an on/off ratio of 10$^{4}$.}
\end{figure*}

Our devices each contain a lightly boron--doped (p--type) mesa--etched single crystal Si nanowire, n$^{+}$--type source and drain, and two layers of gates; see Fig. \ref{sample} (a). The topmost gate layer, which we call the upper gate (UG), covers the entire device between the heavily doped source and drain. Applying a positive voltage to the upper gate inverts the underlying Si nanowire and provides conduction. The second gate layer, which we call the lower gates (LG), consists of three finger gates which wrap around the Si nanowire. These are denoted as LGS (closest to the source), LGC (center gate) and LGD (closest to the drain); see Fig. \ref{sample} (b). The lower gate fingers are primarily used to locally deplete the electron gas and therefore to create electrostatically controlled tunnel barriers (LGS and LGD), or to modulate the electrostatic potential of a quantum dot (LGC). 

The devices are fabricated on a 6 inch silicon--on--insulator (SOI) wafer, with doping density of about 10$^{15}$ cm$^{-3}$, an initial SOI thickness of 100 nm, and a buried oxide (BOX) thickness of 200 nm. To minimize the interface trap density at the gate oxide interface of the nanowire\cite{kapila07,gray66}, we fabricate the SOI nanowires at a 45$^{\circ}$ angle with respect to the flat ($<$110$>$) of the wafer in order to obtain a $<$100$>$ crystallographic equivalent orientation on each facet of the nanowire.

As previously mentioned, we fabricate these devices with a fully CMOS process flow developed at the Center for Nanoscale Science and Technology (CNST) nanofabrication user facility at NIST. The fabrication process is presented in Fig. \ref{fabflowchart}. The nanowire, lower gate and upper gate lithography and etching are performed with negative tone electron beam lithography (EBL) using hydrogen silsesquioxane (HSQ) as a resist  
and dry etching in Cl$_2$ chemistry. Source and drain areas located about 10 $\mu$m away from the active device area are implanted with phosphorous at 30 keV with a dose of 10$^{15}$ cm$^{-2}$. We grow sacrificial oxide on both the nanowire and the lower gate layer in order to remove possible etch damage produced during the dry etch. Both sacrificial oxide and the gate oxide on the nanowire are grown in a tube furnace at 850 $^{\circ}$C and 950 $^{\circ}$C, respectively. The sacrificial oxide is removed with a short 100:1 HF dip. The lower and upper gate layers are 75 nm thick {\it in situ} phosphorous--doped poly--Si deposited by low pressure chemical vapor deposition (LPCVD) at 625 $^{\circ}$C. Both gate layers are degenerately doped to ensure electrical conduction at low temperatures with a typical resistivity of 10--30 m$\Omega\cdot$cm (determined from two terminal measurement at 2.2 K). The sacrificial oxide on the lower gate and the isolation oxide between the lower gate and the upper gate are grown with rapid thermal oxidation (RTO) at 1000 $^{\circ}$C. The final steps of the process are metallization of ohmic contacts with sputter deposited Al--1\%Si and a forming gas anneal at  425 $^{\circ}$C for 30 min.

To date, we have fabricated devices on two 6 inch wafers which we call A and B\cite{Note1}; see Fig. \ref{fabflowchart}. The main differences between the wafers are  the nominal gate oxide thickness and the finger gate lengths. The SOI nanowire width and length are 70 nm and 800 nm respectively for both wafers. Each wafer contained 48 dies: 36 with two devices as in Fig. \ref{sample} on each and 12 diagnostics dies located on the diagonals of the wafer. The diagnostics dies contained conventional field--effect transistors (FET) with a 70 nm wide SOI nanowire as a channel, and test structures to measure the resistance of ohmic contacts and the resistivity of the poly--Si.

A cross-sectional SEM image produced by a focused ion beam (FIB) cut along the LGC finger (white dashed line in Fig. \ref{sample}(b)) of a finished device is shown in Fig \ref{sample}(c). The darker areas in the micrograph are Si and the gray areas are SiO$_{2}$. The cross--sectional image shows that both poly--Si films of upper gate and lower gate layers conformally coat the layers underneath as is expected from LPCVD growth, and that the oxides are continuous, as needed for electrical isolation.  

\section{Results and discussion}
We characterized many devices and FETs from randomly chosen dies across both wafers at room temperature and at 2.2 K. In addition, one of the devices was cooled down and measured in a dilution refrigerator to 30 mK. The summary of the electrical characterization is presented in table \ref{samplesummary}.

Figure \ref{3Gchar} (a) shows turn--on characteristics of different devices (as in Fig. \ref{sample}) on wafer A (40 nm gate oxide). The solid lines and dashed lines correspond to data taken at 300 K and 2.2 K, respectively. At room temperature the threshold voltages, as obtained by linearly extrapolating the current to zero\cite{sze}, were $V_{\text T}=-0.3$ V with the standard deviation of 0.1 V and there was a threshold shift of about 0.6 V when devices were cooled down. A simple estimate of the threshold voltage\cite{schroder} which ignores the presence of any fixed oxide charge and which treats the devices as planar FETs yields $V_{\text T}=-0.1$ V at room temperature and $V_{\text T}=0.4$ V at 2.2 K, which is in reasonable agreement with our measured values. Wafer B (25 nm gate oxide) showed uniform turn--on characteristics with $V_{\text T}\approx0$ V at room temperature and a nearly equal shift in the threshold when cooled down. The diagnostic transistors on each wafer also showed similar turn--on behavior. A typical subthreshold slope for these devices was 80 mV/decade at room temperature (the ideal subthreshold slope is 60 mV/decade\cite{schroder}) with an on--off ratio of 10$^{4}$, see inset in Fig. \ref{3Gchar} (a). 
Typical turn--off characteristics for each of the finger gates (LGS, LGC and LGD) measured at both room temperature and  2.2 K for wafer B are shown in Fig. \ref{3Gchar} (b). The room temperature data was taken with an upper gate voltage $V_{\text UG}=1.3$ V, while the low temperature data was taken with $V_{\text UG}=2$ V. 
The range for turn--off voltages, i.e. the lower gate voltage $V_{\text LG}$ at 100 pA of drain current $I_{D}$, was from -1.5 V to -1 V at room temperature and -1 V to -0.5 V at 2.2 K for all measured lower gates for all devices. 

\begin{figure}[t!]
  \includegraphics[width=\columnwidth]{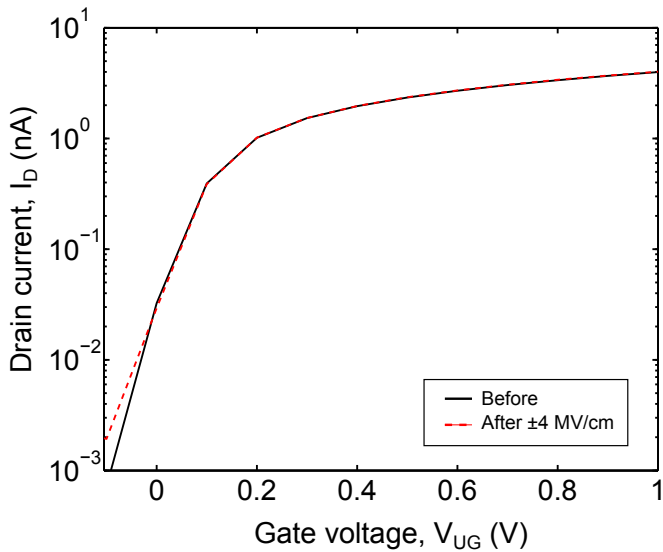}
  \caption{\label{transistor}(Color online) Turn--on characteristics, i.e. drain current as a function of gate voltage of a MOSFET transistor with 25 nm gate oxide thickness. Data was taken with the drain voltage $V_{D}$ of 1 mV. Solid (black) line is initial turn--on curve, dashed (red, gray) line is turn--on curve after $\pm$10 V (4 MV/cm) excursion in the gate voltage. The indentical curves indicate immunity to dielectric breakdown up to $\pm$4 MV/cm. }
\end{figure}

We also tested the robustness of the gate oxide and the isolation oxide on wafer B. In these tests, the gate voltage was swept in steps up to $\pm$10 V while the source-drain and leakage currents were simultaneously measured to the other gates and to the channel. All leakage resistances between the channel and either layer of gates or between gates were $>$10 G$\Omega$ up to gate voltages of $\pm$10 V. After each gate voltage excursion, the turn--on characteristics were remeasured in order to determine if there was a change in the threshold voltage or slope. Diagnostic FETs were immune to electric fields up to 4 MV/cm ($\pm10$ V), showing no change in $V_T$ nor generation of a leakage path (Fig. \ref{transistor}). Similar robustness measurements for SET devices showed no threshold shift up to 2.8 MV/cm ($\pm$7 V) and only a small (0.05 V and 0.2 V) threshold shift after a gate excursion of 4 MV/cm ($\pm$10 V) in two out of four devices. No observable leakage path developed during the sweep. A typical literature value of the breakdown field of metal--oxide--semiconductor capacitor (MOSCAP) is about 10 MV/cm, before generating a leakage path\cite{harari77}. We also performed robustness measurements of the isolation oxide between LG and UG on about 50 different lower gate fingers on different devices across the wafer. Only three fingers developed a breakdown path during the $\pm$10 V sweep. 

\begin{table*}
\scriptsize
\caption{\label{samplesummary}Summary of the wafer characteristics.}
	\begin{tabular}{|c|c|c|c|c|}
	\hline
		Test 													& Wafer A result 		& \# tested & Wafer B result	& \# tested \\
		\hline	
		Threshold voltage value  			& [-0.4 V, -0.1 V] 		& 8					& [-0.1 V, 0.3 V]	& 4 \\
		and uniformity	at 300 K			&										&						&									&			\\
		\hline
		Threshold voltage value 			& [0.2 V, 0.4 V]		& 3					& [0.5 V, 1 V]		& 4 \\
		and uniformity	at 2.2 K 			&										&						&									&	 \\
		\hline
		LG Turn--off voltage value and&  --								& 	--			& [-1.5 V, -1 V] 	& 4 \\
		uniformity	at 300 K					&										&						&								 	& \\
		\hline
		LG Turn--off voltage value 		&  --								& 	--			& [-1 V, -0.5 V] 	& 4 \\
		and uniformity	at 2.2 K 			&										&						&									&		\\
		\hline
		On/off ratio									&	$10^3$													&		8				&	$10^4$											&	4		\\
		\hline
		Subthreshold slope (300 K)		&	--															&		--			&	80 mV/decade 								& 4		\\
		\hline
		UG--channel leakage \footnotemark[1] &	$>$10 G$\Omega$ at $\pm$ 0.25 MV/cm			&	7/8\footnotemark[4]					&	$>$10 G$\Omega$ at $\pm$ 4 MV/cm		& 4/4\footnotemark[4]		\\
		UG breakdown \footnotemark[2]	& no breakdown at $\pm$ 2.5 MV/cm 	&	2/2\footnotemark[4]					&	2 breakdown at $\pm$ 4 MV/cm 	&	2/4\footnotemark[4]		\\
		\hline
		LG--UG leakage								&	$>$10 G$\Omega$ at $\pm$ 0.4 MV/cm			&	24/24\footnotemark[4]				& $>$10 G$\Omega$ at $\pm$ 4 MV/cm 		& 46/49\footnotemark[4]\\
		LG breakdown \footnotemark[3]	& no breakdown at $\pm$ 4 MV/cm		& 6/6\footnotemark[4]					&	3 breakdown at $\pm$ 4 MV/cm& 3/49\footnotemark[4]			\\
		\hline
		Functional nanowires					&	8																&	8					&	4														&	34		\\
		Functional LG fingers 				&	0																&	24				&	12														&	12		\\
		\hline
	\end{tabular}
	\footnotetext{1 sample on wafer A showed leakage resistance of 100 M$\Omega$ to the channel}
	\footnotetext[2]{Breakdown generated threshold voltage shift after the voltage excursion.}
	\footnotetext[3]{Breakdown generated leakage path.}
	\footnotetext[4]{\# tested that showed the result/total \# tested}
\end{table*}
\normalsize
\begin{figure}[t!]
  \includegraphics[width=\columnwidth]{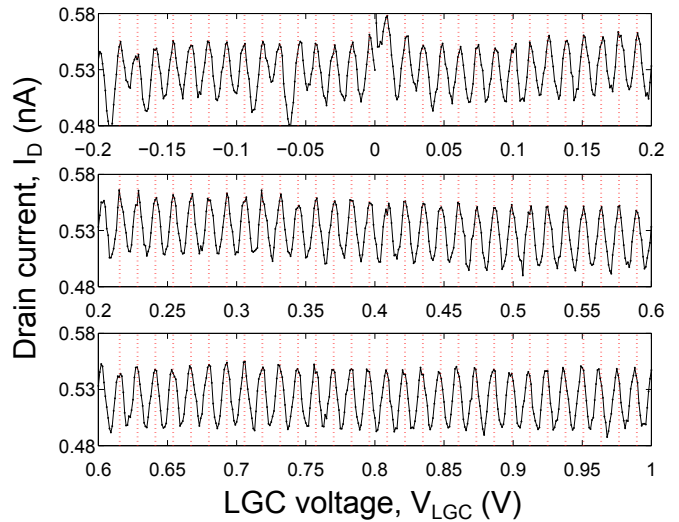}
  \caption{\label{SETosc}(Color online) SET oscillations of a device at 2.2 K, vertical dashed lines (red, gray) are separated by a period of 12.9 mV. The data show good uniformity of the gate capacitance over 90 periods.}
\end{figure}

\begin{figure}[t!]
  \includegraphics[width=\columnwidth]{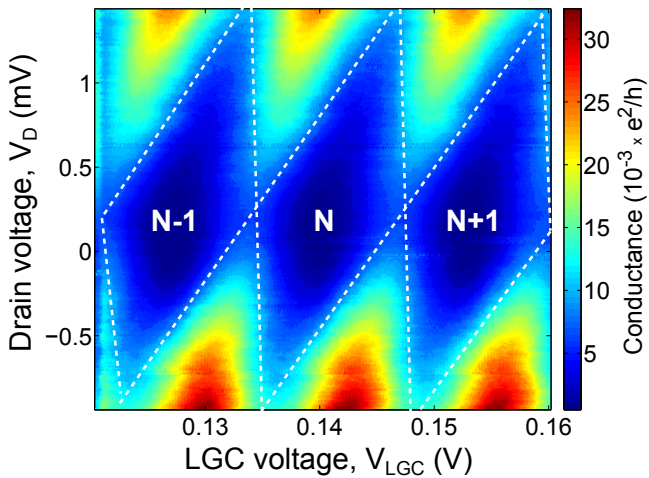}
  \caption{\label{diamonds}(Color online) Coulomb diamonds measured at 30 mK for an SET device, N is the electron number. Charging energy $E_{C}=$1.2 meV, drain capacitance $C_{D}$=7 aF and total capacitance of the island $C_{\Sigma}$=70 aF, extracted from the diamond data. }%
\end{figure}

Above, we have discussed the robustness and uniformity of devices in terms of MOSFET performance, and in the following we present device characteristics when operated in a single electron device mode. First, we discuss ease of tuning. The right hand side of Fig. \ref{sample} (b) shows a schematic of a typical measurement circuit for a device. Tuning the device to display SET oscillations took very little time, on the order of minutes, because there was very little cross capacitance between gates and each barrier was controlled by a single gate voltage. To tune a device into SET mode, we first applied a small  bias voltage to the drain (of order 1 mV) and set the upper gate to a voltage (obtained from a short upper gate sweep, typically about 2 V) which gave about 1 nA of current. Next, a two dimensional sweep of $V_{\text{LGS}}$ and $V_{\text{LGD}}$ (with $V_{\text{LGC}}$ well above the turn--off voltage) was performed to find the voltages where each of these gates began to turn off conduction. Typically, the barrier voltages were about -0.6 V. We note that barrier resistances responded symmetrically to $V_{\text{LGS}}$ and $V_{\text{LGD}}$. After fine tuning $V_{\text{LGS}}$ and $V_{\text{LGD}}$, we measured SET oscillations by sweeping $V_{\text{LGC}}$ with the other gate voltages fixed.     

Coulomb oscillations of a device taken at 2.2 K are presented in Fig. \ref{SETosc}. The oscillations were very regular with period of 12.9 mV over an LGC gate voltage range of 1.2 V, corresponding to about 90 periods. The capacitance of LGC to the dot, extracted from the SET oscillation period, was about 12 aF. Coulomb diamond data recorded at 30 mK for the same device is shown in Fig. \ref{diamonds}. The charging energy and lever arm $\alpha$, which converts the gate voltage to the electrostatic potential of the island $U_{dot}=\alpha V_{LGC}$, extracted from the diamond data were 1.2 meV and 0.09, respectively. The capacitance between the dot and each of the other two lower gates (LGS and LGD) was measured relative to the LGC capacitance by following the position of a current peak when sweeping LGS (or LGD) and LGC (data not shown). The capacitance values for both LGS and LGD were about 5 aF, indicating the dot was located in the center of the device. This, together with the agreement between the measured capacitance to LGC (12 aF) and that calculated from the geometry (14 aF) gives us confidence that the dot being modulated was an intentional dot formed through electrostatic control of LGS and LGD rather than through barriers formed by defects. 

\begin{figure}
  \includegraphics[width=\columnwidth]{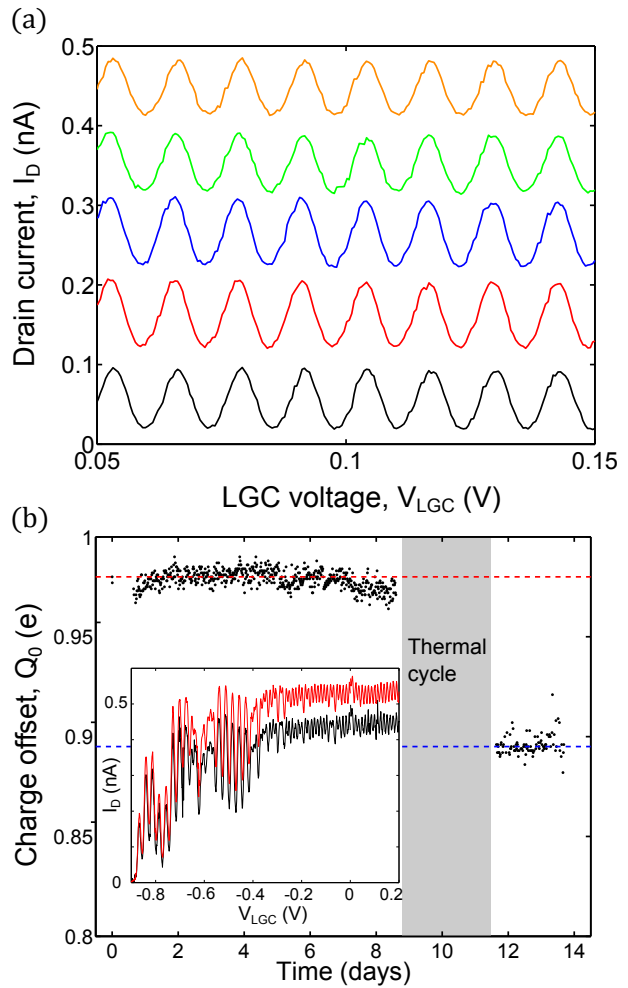}
  \caption{\label{CBdata}(Color online) (a) An example of single electron oscillations taken at different times with interval of about 3 hours; curves are offset vertically for clarity.  (b) Charge offset drift derived from Coulomb oscillations, red and blue (gray) dashed horizontal lines are a guide for the eye. The gray area indicates the time interval of the thermal cycle. After the thermal cycle the charge offset value $Q_{0}$ changed by 0.1 e, but remained as stable as before. Inset: Drain current as a function of LGC voltage $V_{LGC}$ before (black line) and after (red, gray line) the thermal cycle. The data in the inset before and after the thermal cycle are taken with the same gate voltages. All features in the data are reproduced after the thermal cycle.}%
\end{figure}

As a more strict test of the quality of our fabrication we performed charge offset drift measurements on several devices. This measurement consisted of repeatedly measured Coulomb blockade oscillations at a fixed time interval over several days. Figure \ref{CBdata} (a) shows a typical collection of SET oscillations taken at 2.2 K and spanning the total duration of the charge offset drift measurement. To obtain charge offset drift values for each curve, a sinusoidal function of the form $I_{\text{drain}}=I_{0}\sin[2\pi(V/\Delta V+Q_{0}/e)]$ was fit to the measured data. Here $I_{0}$ is the amplitude of the oscillations, $V$ is the gate voltage and $\Delta V$ the oscillation period. The phase of the sinusoidal fit function, $Q_{0}$, is a charge offset value for each curve. The result of this procedure is shown in Fig. \ref{CBdata} (b). The devices exhibited very stable behavior with a drift in $|\Delta Q_0(t)|\approx0.01$ e over 8 days of measurement. Moreover, these results rival those of similar Si devices fabricated in other foundries\cite{NeilJAP,zimmerman07a,hourdakis08a}. In addition, after eight days of measurement, we thermally cycled the device to room temperature. Figure \ref{CBdata} shows this data as well. While the thermal cycle resulted in a shift in the charge offset value of about (0.1$\pm n$) e, the level of drift observed was identical both before and after the thermal cycle. Finally, a measurement of $I_{\text{D}}$ vs. $V_{\text{LGC}}$ voltage with the same LGS, LGD, and UG voltages after the thermal cycle not only reproduced the SET oscillations at the same value $V_{\text{LGC}}$ with charge offset of 0.1 e, but also reproduced the aperiodic features which become prominent when the gate begins to turn off conduction (inset of Fig. \ref{CBdata}).

While the previous results indicate that the cleanliness of our CMOS fabrication is quite good, many devices in this first device run failed electrically by either not turning on or through an inability to turn off conduction with the lower gate fingers. This drove our yield of fully functioning (in which we were able to measure intentional Coulomb blockade) devices down to 4/34 devices measured. We have been able to identify the gross fabrication failures, by cross-sectioning devices with FIB in conjunction with the electrical results. In brief, the failure modes are the result of over--oxidation of the SOI nanowire and the LG fingers, as well as the overall amount of oxide removed in the processing. We believe further development of our process flow will ameliorate these failures.

\section{Summary and Conclusions}

In summary, we have demonstrated devices which show good uniformity in electrical
characteristics from device to device within a wafer and between wafers. Moreover, the
devices are quite robust against dielectric breakdown up to electric fields of 4
MV/cm. Finally, and most importantly, when operated as a single electron device, these
devices show very stable behavior. Taken together, these characteristics indicate a
relatively clean and stable electrostatic environment throughout the fabrication
process. We attribute these successes to the minimization of impurities and defects which
result from our CMOS processing and material restrictions. To further improve the usefulness of these 
devices as current standards and quantum information devices, our next tasks include i)
substantially increasing the yield, and ii) making shorter finger gates so that we can use
those gates to both generate barriers and as plunger gates.

While these results indicate that a fully CMOS process pays dividends in device performance,
it also complicates the fabrication. We believe that our results in terms of reliability,
ease of tuning, and clean SET behavior all justify the cost of the increased complexity of
fabrication.


%
%
%

\begin{acknowledgments}
  We thank Ted Thorbeck, John Bonevich, Jerry Bowser, and Vincent Luciani for fruitful discussions. We also thank Jerry Bowser and Vincent Luciani for guidance with the fabrication. Research was performed in part at the NIST Center for Nanoscale Science and Technology (CNST).
\end{acknowledgments}

\providecommand{\noopsort}[1]{}\providecommand{\singleletter}[1]{#1}%
%


\end{document}